\documentclass[
  aps, 
 superscriptaddress,
 prl,
 reprint,
 showpacs,
 amsmath,amssymb,
]{revtex4-1}

\usepackage{graphicx}
\usepackage{color}
\usepackage{dcolumn}

\usepackage{bm}

\usepackage{xcolor}
\usepackage{color}
\usepackage{hyperref} 
\hypersetup{
    urlbordercolor=black, 
  colorlinks   = true, 
  urlcolor     = blue, 
  linkcolor    = blue, 
  citecolor   = blue 
}

\begin{document}

\preprint{APS/123-QED}

\title{Wake-mediated propulsion of an upstream particle in two-dimensional plasma crystals}

\author{I.\ Laut}
\email{Ingo.Laut@dlr.de}
\affiliation{Institut f\"ur Materialphysik im Weltraum, Deutsches Zentrum f\"ur Luft- und Raumfahrt (DLR), 82234 We{\ss}ling, Germany}

\author{C.\ R\"{a}th}
\affiliation{Institut f\"ur Materialphysik im Weltraum, Deutsches Zentrum f\"ur Luft- und Raumfahrt (DLR), 82234 We{\ss}ling, Germany}

\author{S.\ K.\ Zhdanov}
\affiliation{Institut f\"ur Materialphysik im Weltraum, Deutsches Zentrum f\"ur Luft- und Raumfahrt (DLR), 82234 We{\ss}ling, Germany}

\author{V.\ Nosenko} 
\affiliation{Institut f\"ur Materialphysik im Weltraum, Deutsches Zentrum f\"ur Luft- und Raumfahrt (DLR), 82234 We{\ss}ling, Germany}

\author{G.\ E.\ Morfill}
\affiliation{Max Planck Institute for Extraterrestrial Physics, 85741 Garching, Germany}
\affiliation{BMSTU Centre for Plasma Science and Technology, Moscow, Russia}

\author{H.\ M.\ Thomas}
\affiliation{Institut f\"ur Materialphysik im Weltraum, Deutsches Zentrum f\"ur Luft- und Raumfahrt (DLR), 82234 We{\ss}ling, Germany}

\date{\today}

\begin{abstract}
{
The wake-mediated propulsion of an ``extra'' particle in a channel of two neighboring rows of a two-dimensional plasma crystal, observed experimentally by Du \emph{et al.} [Phys. Rev. E {\bf 89}, 021101(R) (2014)], is explained in simulations and theory. We use the simple model of a pointlike ion wake charge to reproduce this intriguing effect in simulations, allowing for a detailed investigation and a deeper understanding of the underlying dynamics. We show that the nonreciprocity of the particle interaction, owing to the wake charges, is responsible for a broken symmetry of the channel that enables a persistent self-propelled motion of the extra particle. 
We find good agreement of the terminal extra-particle velocity with our theoretical considerations and with experiments.
}
\end{abstract}


\maketitle

\emph{Introduction.}
In ion beam physics, channeling effects can strongly influence the motion of ions or other charged particles in a crystalline solid \cite{gemmell1974, feldman2012}. Similarly, neutral atoms can be channeled in a standing wave of laser light \cite{salomon1987, keller1999}.
In complex plasma crystals, experimental observations of the channeling effect showed that instead of slowing down, an extra particle \emph{accelerates} in the channel \cite{du2012, du2014, zhdanov2016}. This persistent motion was attributed to the nonreciprocity of the particle interactions but the exact origin of the propulsion was not resolved \cite{du2012}. (Self-)propelled motion currently receives considerable attention both in macroscopic \cite{deseigne2010} and microscopic \cite{schaller2010, buttinoni2013} systems.

Complex plasmas consist of micron-sized particles that are immersed in a weakly ionized gas. In a laboratory radio-frequency (rf) plasma, the particles are usually negatively charged and levitate in the plasma sheath region above the lower electrode where the gravitational force is balanced by the electric field. Thus confined, these strongly coupled systems can form two-dimensional (2D) crystalline structures which are called \emph{plasma crystals} \cite{chu1994, *thomas1994, *hayashi1994, morfill2009}. Large three-dimensional crystals can only be obtained under microgravity conditions, for example during parabolic flights \cite{piel2006} or onboard the International Space Station \cite{thomas2008}. 
Both two- and three-dimensional plasma crystals 
are ideal model systems for phase transitions \cite{schweigert1998, killer2016}, wave processes \cite{nunomura2005wave, tsai2016} and self-organization \cite{menzel2010, williams2014}, as their dynamics can be resolved at the level of individual particles.

The sheath electric field not only levitates the crystal, but also causes an ion flow that strongly influences the particle interaction. In the bulk plasma, far away from the rf electrodes, the particle interaction is well described by a screened Coulomb (Yukawa) potential since the charged particles are surrounded by a cloud of positively charged ions \cite{ikezi1986}. In the sheath region, however, the downward-flowing ions distort the screening cloud, leading to a positive excess charge below each particle. This \emph{ion wake} adds an attractive component to the mutual particle interactions \cite{melzer1999} and makes them nonreciprocal. Ion wakes cause interesting effects like the formation of particle strings in a vertically extended system \cite{kong2011}, the mode-coupling instability  in a monolayer \cite{couedel2011} and the coexistence of two distinct kinetic temperatures in a binary mixture \cite{ivlev2015statistical}. A common way to model the ion wake is by a positive pointlike charge that is positioned a fixed distance below each particle. The intuitive picture of a pointlike wake charge allows for the rigorous analysis of the mode-coupling instability \cite{ivlev2000} and of the slightly bowl-like shape \cite{rocker2014wakeinduced} of plasma crystal monolayers.

It was suggested that the ion wake be also responsible for the channeling effect in plasma crystals that was first observed in experiments by Du \emph{et al.} \cite{du2012}, but the exact driving mechanism was not known. An ``extra'' particle floated slightly above the plasma crystal (\emph{upstream} with respect to the ion flow) and followed the channel formed by lines of neighboring particles. Despite ambient gas friction, the particle moved at a nearly constant velocity, provoking lateral waves and an increase of kinetic temperature in the crystal \cite{du2012, du2014}. 

In this Letter, we reproduce in simulations the wake-mediated propulsion of an upstream extra particle in a 2D plasma crystal. We use the simple model of a pointlike ion wake charge which enables an intuitive picture and an analytical analysis of the underlying dynamics. The attraction between the extra-particle wake and the particles in the crystal results in a symmetry-breaking deformation of the channel which accelerates the extra particle. We study the terminal velocity reached by the propulsion process and compare it to our theoretical considerations and to experiments.

\emph{Simulation particulars.}
Molecular-dynamics (md) simulations are well suited for modeling the dynamical effects in complex plasmas \cite{totsuji2001, ivlev2003, sheridan2008, laut2016anisotropic}. In the simulation of a 2D plasma crystal in the horizontal $xy$ plane, the equation of motion for particle $i$ reads:
\begin{equation}\label{eq_EqMotion}
M\ddot{\mathbf{r}}_i + M \nu \dot{\mathbf{r}}_i = \sum_{j \neq i} \mathbf{F}_{ji} + \mathbf{C}_i^{p,t} + \mathbf{L}_i,
\end{equation}
where $\mathbf{r}_i$ is the three-dimensional particle position, $M$ the mass and $\nu $ the damping rate. The forces acting on the particle are the mutual particle interactions $\mathbf{F}_{ji}$, the confinement force $\mathbf{C}_i^{p, t}$ derived from an external potential, and a Langevin heat bath $\mathbf{L}_i$. 

To include the ion wake in the mutual particle interaction, a positive pointlike charge $q$ is placed a fixed vertical distance $\delta$ below each particle, while the particle itself is modeled as a negative pointlike charge $Q < 0$. The force exerted by particle $j$ (and its wake) on particle $i$ is thus modeled as
\begin{equation}
\label{eq_interparticel_forces}
\begin{split}
\mathbf{F}_{ji} & =                                     
 Q^2 f( r_{ji} ) \frac{\mathbf{r}_{ji}}{r_{ji}}
 +qQ f( r_{ w_{ji} } ) \frac{ \mathbf{r}_{ w_{ji} } }{ r_{ w_{ji} } } \\ 
 & \equiv F_p \frac{\mathbf{r}_{ji}}{r_{ji}}  + F_w \frac{\mathbf{r}_{w_{ji}}}{r_{w_{ji}}} , \\                
\end{split}
\end{equation}
where $f(r) = \exp ( -r/\lambda) ( 1 + r/\lambda) / r^2$, $\lambda$ the screening length, 
$\mathbf{r}_{ji} = \mathbf{r}_{i} - \mathbf{r}_{j}$ and 
$\mathbf{r}_{w_{ji}} = \mathbf{r}_{i} - (\mathbf{r}_{j} - \delta \mathbf{e}_z)$. Here and in the following, $r$ denotes the magnitude of vector $\mathbf{r}$, and $\mathbf{e}_{x,y,z}$ are the unit vectors of the coordinate system. 

The confinement force reads $\mathbf{C}_i^{p, t} = -\bm{\nabla} (V_i^{p, t} + V_{i}^z)$, where $V_i^{p, t}$ is the horizontal confinement and $V_i^z$ the vertical confinement. Very regular crystals with few defects are obtained with a horizontal tenth-order potential $V_i^t = 0.5 M \Omega_h^2 \rho_i^{10}/R^8$, where $\rho_i = \sqrt{x_i^2 + y_i^2}$ is the horizontal position of particle $i$ and $R$ is approximately the radius of a crystal with the same number of particles in a parabolic horizontal confinement with frequency $\Omega_h$ \cite{durniak2010}. 
The parabolic horizontal potential $V_i^p = 0.5 M \Omega_h^2 \rho_i^2$ leads to a crystal where the interparticle distance increases with the distance from the crystal center \cite{totsuji2001, ivlev2003, laut2016anisotropic}. While the global structure of a plasma crystal is well described by a parabolic confinement \cite{zhdanov2003large}, the tenth-order potential is well suited to reproduce the regular central region.

The vertical confinement of the particles stems from the interplay of the gravitational force and the electrostatic forces of the sheath field which are oriented in opposite directions. This strong confinement is often modeled as a parabolic confinement $V_i^z = 0.5 M \Omega_z^2 (z_i - z_\text{eq})^2$, where the equilibrium position is $z_\text{eq}=0$ \cite{totsuji2001, ivlev2003, rocker2014nonlinear}. In order to reproduce an upstream particle in simulations, we used a different equilibrium position $z_\text{eq} = h$ for the extra particle than for the other particles that were confined at $z_\text{eq} = 0$.

The Langevin force $\mathbf{L}_i(t)$ is defined by $\langle \mathbf{L}_i(t) \rangle = 0$ and $\langle \mathbf{L}_i(t + \tau) \mathbf{L}_j(t) \rangle = 2 \nu M T \delta_{ij} \delta(\tau)$, where $T$ is the temperature of the heat bath, $\delta(t)$ the delta function and $\delta_{ij}$ the Kronecker delta.

\begin{figure}
\includegraphics[width=\columnwidth]{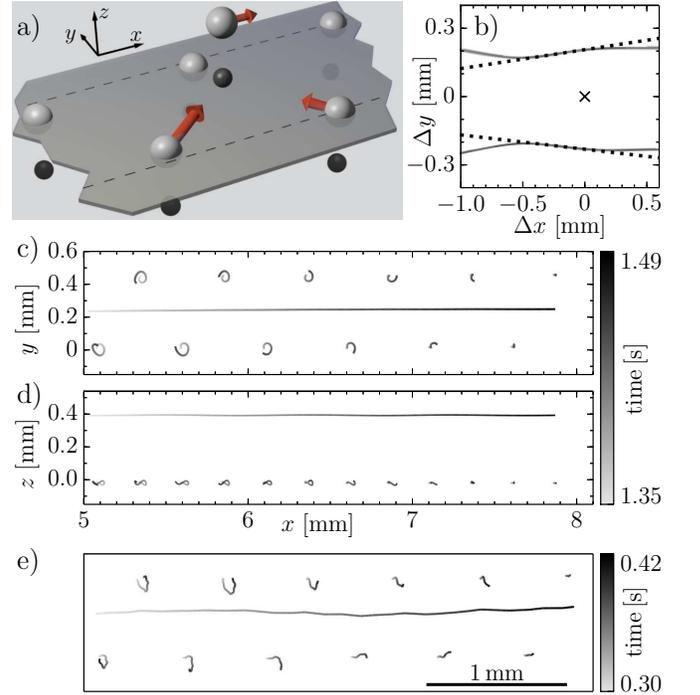}
\caption{%
(Color online) Channeling of an ``extra'' particle in a simulated complex plasma crystal. 
(a) The particles are sketched in gray, their wake charges in black. The unperturbed channel is visualized by dashed lines on the semi-transparent crystal plane. During the passage of the extra particle floating above the crystal from left to right, its wake attracts the channel particles (thick arrows shown for two particles). Due to the small displacement $\delta y$ of the channel particles, the repulsive particle interaction is stronger behind the extra particle than in front of it, leading to a propulsion of the extra particle (thin arrow).
(b) Horizontal particle positions with respect to the extra particle between $t_1 = 1.35$~s and $t_2 = 1.49$~s. The position of the extra particle is indicated by a cross. The relative positions were fitted by straight dotted lines for $-0.4~\text{mm} < \Delta x < 0$.
(c) Horizontal (top view) particle positions color-coded with respect to time between $t_1$ and $t_2$. 
(d) The same for the $xz$ projection (side view); here, only the extra particle and the two neighboring rows which form the channel are shown.
(e) Horizontal particle positions from Experiment~5 of Ref.~\cite{du2012}.
} \label{fig_1_depict_channeling}
\end{figure}

\begin{figure}
\includegraphics[width=\columnwidth]{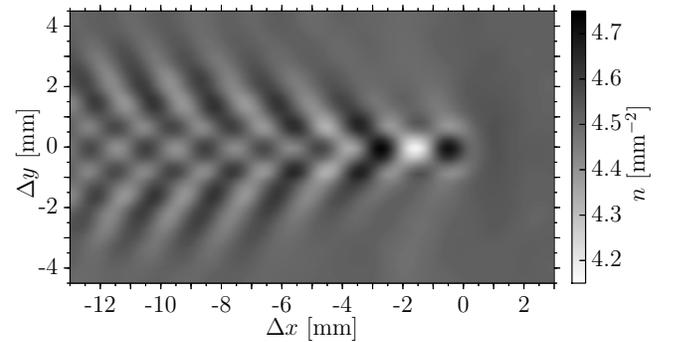}
\caption{%
Local particle density $n$ in the reference frame of the extra particle, showing the pronounced wave structure behind the extra particle (lateral wake \cite{dubin2000, nosenko2003}).
}
\label{fig_2_subsonic_lateral_wave}
\end{figure}

In a simulation run, $N-1$ particles are initially equilibrated. The $x$ and $y$ axes are oriented so that there is a line of nearest neighbors (and a channel) in the $x$ direction, see Fig.~\ref{fig_1_depict_channeling}(c). Then, at $t=0$, the extra particle is added to the crystal with an initial velocity $\mathbf{v}(t=0) = v_0 \mathbf{e}_x$ in the $x$ direction. The simulation is stopped before the extra particle reaches the boundary of the crystal.

The simulated crystal consisted of $N=10001$ particles, the particle mass $M = 0.61 \times 10^{-12}$~kg, charge $Q=-18500e$ and screening length $\lambda = 380$ $\mu$m was in the parameter range of the experimental observations of Refs.~\cite{du2012, du2014}. If not stated otherwise, the friction coefficient was $\nu = 1.26$~s$^{-1}$, corresponding to a typical gas pressure of 1.0~Pa \cite{liu2003}. The wake charge $q$ and distance $\delta$ are not known in experiments---in theory and simulations they are assumed to be a fraction $<1$ of the particle charge and of the screening length, respectively \cite{ivlev2000, rocker2014nonlinear, laut2016anisotropic}. Here, $q = 0.6|Q|$ and $\delta = 0.4 \lambda$ was used. The parameters of the confinement were $\Omega_h = 2\pi \cdot 0.12~\text{s}^{-1}$, $\Omega_z = 2\pi \cdot 22~\text{s}^{-1}$, $h = 350~\mu\text{m}$ and $R = 27$~mm.

\emph{Channeling effect.}
The horizontal trajectory of the extra particle between $t_1 = 1.35~\text{s}$ and $t_2 = 1.49~\text{s}$ can be seen in Fig.~\ref{fig_1_depict_channeling}(c). The extra particle was added to the crystal at the position $\mathbf{r}_e(t=0) = (-18~\text{mm}, 0.21~\text{mm}, 0.39~\text{mm})$ with an initial velocity $v_0 = 10~\text{mm/s}$. The particles were confined by the tenth-order potential $V_i^t$ and formed a very regular lattice with an interparticle distance of $a = 504 \pm 3~\mu\text{m}$. It can be seen that the particles forming the channel (the \emph{channel particles} in the following) move towards the extra particle shortly after its passage. They subsequently perform a circular motion. Although a friction force is applied to the extra particle, it moves in a straight line at a constant velocity. The horizontal trajectories agree well with the experimental data shown in Fig.~\ref{fig_1_depict_channeling}(e). In Fig.~\ref{fig_1_depict_channeling}(d), it can be seen that the channel particles oscillate also in the vertical direction, while the extra particle moves at an almost constant height. 

To visualize the deformation of the channel, the horizontal particle positions with respect to the extra particle, $\Delta \mathbf{r}_i = \mathbf{r}_i - \mathbf{r}_e$, are shown in Fig.~\ref{fig_1_depict_channeling}(b). It can be seen that behind the extra particle the channel is clearly deformed, breaking the forward-backward symmetry of the channel geometry. The channel is shaped as a cone in the vicinity of the extra particle. In a range $-0.4~\text{mm} < \Delta x < 0$, the deformed channel is fitted to straight lines which have angles of $9.4^\circ$ and $-7.1^\circ$ measured from the $x$ axis.

The sketch shown in Fig.~\ref{fig_1_depict_channeling}(a) gives a first idea of the propulsion mechanism: During the passage of the extra particle floating above the crystal, the channel particles are attracted to the wake charge of the extra particle and thus deform the channel. Due to this symmetry-breaking deformation, the repulsion between the channel particles and the extra particle is stronger behind the extra particle than in front of it, leading to a net propelling force acting on the extra particle. 

\begin{figure}
\includegraphics[width=\columnwidth]{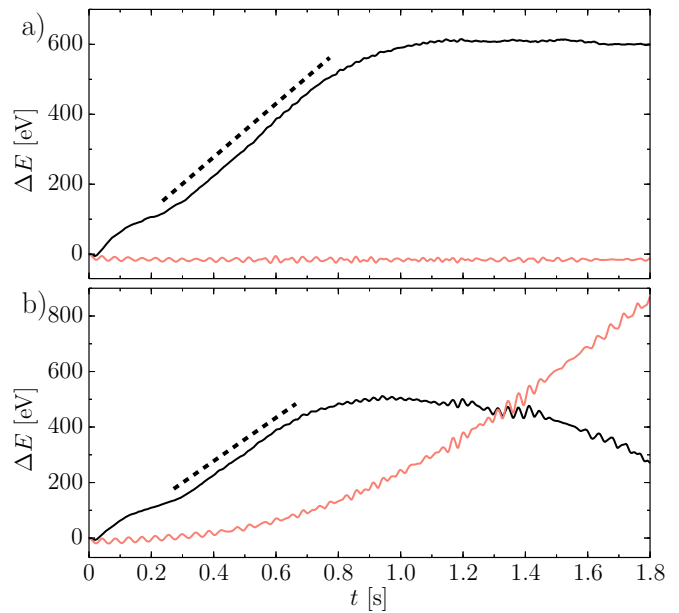}
\caption{%
(Color online) (a) Variation of the kinetic energy (black) and the potential energy (red/gray) of the extra particle as a function of time. The dashed line of slope $765~\text{eV/s}$ is to guide the eye. (b) The same for a crystal confined horizontally by a parabolic confinement. The dashed line has a slope of $781~\text{eV/s}$.
}
\label{fig_3_energy_balance}
\end{figure}

The density variations in the crystal caused by the extra particle are shown in Fig.~\ref{fig_2_subsonic_lateral_wave}. The density in the reference frame of the extra particle was averaged over times $t_1 < t <t_2$. Pronounced subsonic lateral wakes behind the extra particle can be clearly seen. They form due to the dispersion of waves that are excited by the extra particle and propagate in the crystal \cite{dubin2000}. The structure of the density variations is very similar to the experimental observation in Ref.~\cite{du2012}. Just behind the extra particle, in a region $-1~\text{mm} < \Delta x < 0$ and $|\Delta y| < 0.5~\text{mm}$, the density is substantially increased. Similar lateral wakes have been produced in experiment and simulation by sweeping an external perturbation through the crystal \cite{nosenko2003}.

The extra-particle energy balance is depicted in Fig.~\ref{fig_3_energy_balance}(a). The potential energy of the particle, given by the mutual particle interactions of Eq.~(\ref{eq_interparticel_forces}) and the confinement $V_i^t + V_i^z$, does hardly change, it is only modulated by a slight vertical oscillation of the particle. The kinetic energy of the particle increases linearly in the first second of the channeling process, before saturating at an almost constant value. 

In a different simulation, the crystal was confined by the parabolic horizontal confinement $V_i^p$. The extra particle was initially positioned near the center of the crystal with $v_0 = 10~\text{mm/s}$. The extra particle was confined in the channel which was slightly bent (see also Supplemental Material~\footnote{See Supplemental Material at \textcolor{red}{URL} for a movie of an extra particle in a crystal with parabolic confinement.}). The energy balance is shown in Fig.~\ref{fig_3_energy_balance}(b). As the extra particle advances, it accumulates potential energy due to the parabolic confinement, but the propulsion effect is strong enough to accelerate the particle. Only near the boundary of the crystal the velocity decreases as the confinement well becomes steeper.

\begin{figure}[t!]
\includegraphics[width=\columnwidth]{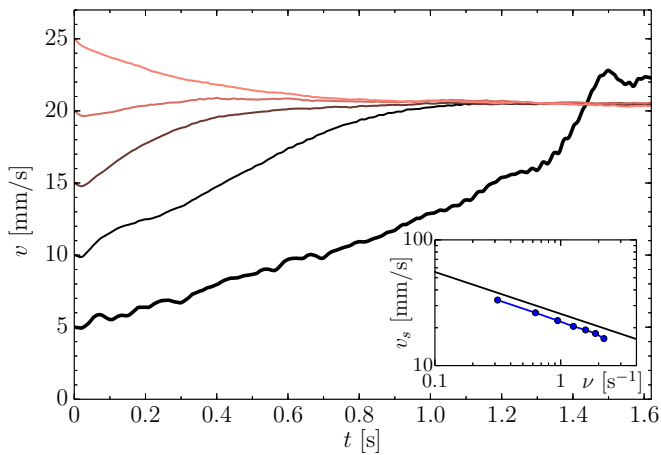}
\caption{%
(Color online) Magnitude of the extra-particle velocity as a function of time, for different initial velocities $v_0$. The extra particle with $v_0 = 5~\mathrm{mm/s}$ (bold line) exits the channel at $t \approx 1.4~\text{s}$ while 
the others stay in the same channel and reach a constant velocity $v_s$. 
The inset shows $v_s$ as a function of damping rate $\nu$. 
The error is no larger than the symbol size. The predicted value of Eq.~(\ref{eq_analytic_model}) is shown as a black line. }
\label{fig_4_final_velocities}
\end{figure}

\emph{Model for accelerated extra-particle motion.} 
In Fig.~\ref{fig_4_final_velocities}, the velocity of the extra particle $v$ in a tenth-order horizontal confinement is shown as a function of time for five simulations with different initial velocities $v_0$ in the interval containing the longitudinal and transverse sound speeds $c_L = 23 \pm 1$~mm/s and $c_T = 6 \pm 1$~mm/s. In the cases where $10~\text{mm/s} \leq v_0 \leq 25~\text{mm/s}$, the extra particle is confined in the channel and its velocity saturates at $v_s = 20.5 \pm 0.3~\text{mm/s}$ where the propulsion effect is counterbalanced by friction. Note that $v_s$ is also reached from above for particles with $v_0 > v_s$. In the simulation with $v_0 = 5~\text{mm/s}$ (bold line in Fig.~\ref{fig_4_final_velocities}), the channel particles are displaced such that the repelling force exerted on the extra particle has a larger $y$ component, leading to a less effective propulsion in $x$ direction. It also leads to a larger scattering angle in the channel, and as the extra particle accumulates enough kinetic energy it leaves the channel at $t \approx 1.4~\text{s}$. Upon exiting the channel, the extra particle is greatly accelerated such that it obtains a velocity $v > v_s$.

Below we establish a simple model for the propulsion mechanism in the channel.
To estimate the small displacement $\delta y$ of the channel particle, we assume the attraction to the wake of the extra particle $F_w(r_w)$ to be constant in the range $-a/2 < \Delta x < a/2$ and evaluate it at $\mathbf{r}_w = (a/2, b/2, h-\delta)$, i.e., when the particle distance in $x$ direction is half an interparticle distance [as sketched in Fig.~\ref{fig_1_depict_channeling}(a)]. This yields $\delta y \approx F_w(r_w) b a^2 / (4 r_w M v_s^2)$, where $b = \frac{\sqrt{3}}{2}a$ is the channel width. The net propelling force---stemming from the asymmetry of the repulsive particle interactions $F_p(r)$ before and after the passage---can then be estimated as $\Delta F \approx F_p(\tilde r) a/2 \tilde r - F_p(r) a / 2 r$, where $\mathbf{r} = (a/2, b/2, h)$ and $ \mathbf{\tilde r} = (a/2, b/2-\delta y, h)$. For small $\delta y$ we obtain $\Delta F \approx \delta y [ \frac{\partial}{\partial \, \delta y} F_p(\tilde r)a/2 \tilde r]_{\delta y = 0}$, and equating $\Delta F$ with the friction force, the terminal extra-particle velocity can be estimated as
\begin{equation}
\label{eq_analytic_model}
v_\text{theory}^3 \approx 
 \xi(r)  
\frac{ F_p(r) }{r^3}  
\frac{ F_w(r_w) }{r_w} 
\frac{ b^2 a^3}{ 16 M^2 \nu},
\end{equation}
where $\xi(r) = (3 + 3 r/\lambda + r^2/\lambda^2)/(1 + r/\lambda)$ depends rather weakly on $r$. For the simulation parameters, $v_\text{theory} = 23.9~\text{mm/s}$ slightly overestimates the terminal extra-particle velocity. One reason for the discrepancy is that the particle-particle repulsion was not considered when estimating the wake-mediated displacement $\delta y$.
A comparison of $v_s$ and $v_\text{theory}$ for different values of friction rate in the range $0.32~\text{s}^{-1} < \nu < 2.21~\text{s}^{-1}$ can be seen in the inset of Fig.~\ref{fig_4_final_velocities}. The model always slightly overestimates the extra-particle velocity, but the predicted scaling $v_s \propto \nu^{-1/3}$ is in good agreement with simulations. 

\emph{Discussion.}
It is expected that the extra particle floats above the crystal because it has a slightly smaller mass than the other particles \cite{zhdanov2015spontaneous}. Varying the mass (and charge) of the extra particle may yield further conclusions regarding the mass ratios in experiments. In order to fully understand the origin of the extra particles floating above or below the crystal layer, a more realistic particle confinement is needed that explicitly considers the balance of electric and gravitational forces in the vertical direction. The ion wake of the extra particle will be dynamically distorted when it is very close to a channel particle. The point-like model may thus not be applicable in narrow channels (when $a$ is very small) or during head-on collisions with a channel particle.

In Ref.~\cite{schweigert2002}, a particle-in-cell simulation was used to model the particle interactions in a crystal with an extra particle below the crystal plane. The propelled motion of this \emph{downstream} particle was reproduced in md simulations, but no intuitive picture was given for the mechanism. Here, with the aid of the pointlike wake charge model, this intuitive explanation was given for the persistent motion of an upstream particle. The propulsion mechanism based on the nonreciprocal particle interactions controls the terminal extra-particle velocity and enables the study of self-propelled motion in complex plasmas \cite{bechinger2016}.

Reproducing the channeling effect in simulations also enables the study of the wave processes that are discussed in the Supplemental Material \footnote{See Supplemental Material at \textcolor{red}{URL}, which includes Refs. \cite{donko2008, couedel2009, nunomura2003, couedel2010, couedel2016}, for a detailed analysis of the wave processes.}.

Equation \ref{eq_analytic_model} closely reproduces the extra-particle velocity $v_s$ measured in experiments. Assuming the wake parameters to be $q/|Q| = 0.6$ and $\delta/\lambda = 0.4$ as in our simulations, we obtain $v_s/v_\text{theory} =  0.85$, $0.85$  and $0.82$ for experiments 1, 3 and 5 of Ref.~\cite{du2012}, respectively (see Table 1 in Ref.~\cite{du2012}). Again, the predicted velocity is slightly above the observed value. In simulations, channeling was observed if the initial velocity $v_0$ was above the transverse sound speed $c_T$. 

To conclude, we reproduced in md simulations the propulsion of an extra particle in a 2D plasma crystal with the intuitive model of a pointlike ion wake charge. The nonreciprocal particle interactions, owing to the ion wake, lead to an asymmetric deformation of the channel and a net propelling force. The terminal velocity reached by the extra particle agrees well with our theoretical considerations and with experiments.

\begin{acknowledgments}
\emph{Acknowledgments.} %
We thank \textsc{Cheng-Ran Du} for providing the experimental data. This work was supported by the German Federal Ministry for Economy and Technology under grant No.~50WM1441. GEM wishes to acknowledge support from RSF Grant No.~14-43-00053.
\end{acknowledgments}

\bibliography{./../../../10_papers/literature}

\end{document}